%% file: main.tex
\documentclass[a4paper,twocolumn]{article}
\usepackage{preamble}
\begin{document}

\title{Optimization study of scintillator shape of electromagnetic calorimeter for Higgs factories}
\author[1]{Takanori~Mogi\thanks{Talk presented at the International Workshop on Future Linear Colliders (LCWS2019), Sendai, Japan, 28 October-1 November, 2019. C19-10-28.}}
\author[2]{Tomohiko~Tanabe}
\author[2]{Wataru~Ootani}
\author[2]{Satoru~Yamashita}
\author[3]{Ryousuke~Shirai}
\author[3]{Tohru~Takeshita}
\author[4]{Yazhou~Niu}
\author[4]{Jianbei~Liu}
\affil[1]{Department of Physics, Graduate School of Science, The University of Tokyo}
\affil[2]{International Center for Elementary Particle Physics, The University of Tokyo}
\affil[3]{Department of Physics, Faculty of Science, Shinshu University}
\affil[4]{Department of Modern Physics, School of Physical Scinences, University of Science and Technology of China}
\date{}

\twocolumn[
  \maketitle
  \begin{abstract}
    \vspace{-2.5mm}
    \input{abstract.tex}
  \end{abstract}
  \vspace{5mm}
]
\saythanks

\input{contents.tex}

\bgroup
  \setlength\bibitemsep{10pt}
  \printbibliography
\egroup

\end{document}

%% file: abstract.tex
Scintillator-based calorimeters for experiments at Higgs factories (e.g.\ ILC) demand scintillator designs that can detect sufficient number of photons and have good light yield uniformity, and that they can be easily mass-produced. In order to meet these requirements, scintillator strips with a small dimple has been proposed. In our study, we measure the light yield of a dimple scintillator sample; we then compare the measurements with light tracing simulation using GEANT4. We intend to use our results to propose an optimized scintillator shape.

%% file: contents.tex
\section{Introduction}
Colliders known as Higgs factories (e.g.\ International Linear Collider (ILC)\cite{ILC_TDR_vol1}) are future electron-positron colliders to search for new physics beyond the standard model. 
Detectors for Higgs factories, such as International Large Detector (ILD)\cite{ILC_TDR_vol4} of ILC, are designed to have an unprecedented jet energy resolution by combining the capabilities of trackers and calorimeters. Particle Flow Algorithm (PFA) is a method to reconstruct particles in a jet that takes advantage of the best measurements available for the type of particle, and it can realize the required energy resolution. According to PFA, $5 \times 5\ \mathrm{mm^2}$ size readout is preferred for the electromagnetic calorimeter (ECAL) of ILD\cite{ILD_LoI}. To achieve this requirement using scintillator-based calorimeter (ScECAL), Strip Splitting Algorithm (SSA) which reconstructs scintillator strips as virtual square tiles is used\cite{KOTERA_SSA}.

ScECAL consists of scintillator strips, wrapped by reflector films, and Silicon Photomultipliers (SiPMs). The size of a scintillator strip under consideration is $5\ \mathrm{mm}\ \times\ 45\ \mathrm{mm}\ \times\ 2\ \mathrm{mm}$, and $10\ \mathrm{\mu m}$ or $15\ \mathrm{\mu m}$ is being considered as the pixel pitch of a SiPM. Using strip-shaped scintillator, the number of readout channels is 1/10 of that for square-shaped scintillator. However, strip scintillator requires sufficient light yield for Minimum Ionising Particle (MIP) and uniform light yield for various incident positions. In order to get the best performance of ScECAL, the strip shape has to be optimize.

Conventional readout methods for ScECAL are side readout and bottom readout. In the side readout method, a SiPM is placed at the $5\ \mathrm{mm}\ \times\ 2\ \mathrm{mm}$ side of a scintillator. The method has good light yield, but the uniformity is bad. The bottom readout method which a SiPM is placed at the $45\ \mathrm{mm}\ \times\ 5\ \mathrm{mm}$ side has good uniformity, but it has less light yield. Recently, a new readout method, dimple readout was proposed. In this method, there is a dimple at the center of the $45\ \mathrm{mm}\ \times\ 5\ \mathrm{mm}$ side of a strip and a SiPM is implanted into the dimple. To evaluate this new method, we explore the characteristics of the dimple readout method through light yield measurement. This result is also used for optimizing the strip shape.

\section{Measurement}
\begin{figure}[H]
  \centering
  \includegraphics[width=0.6\linewidth]{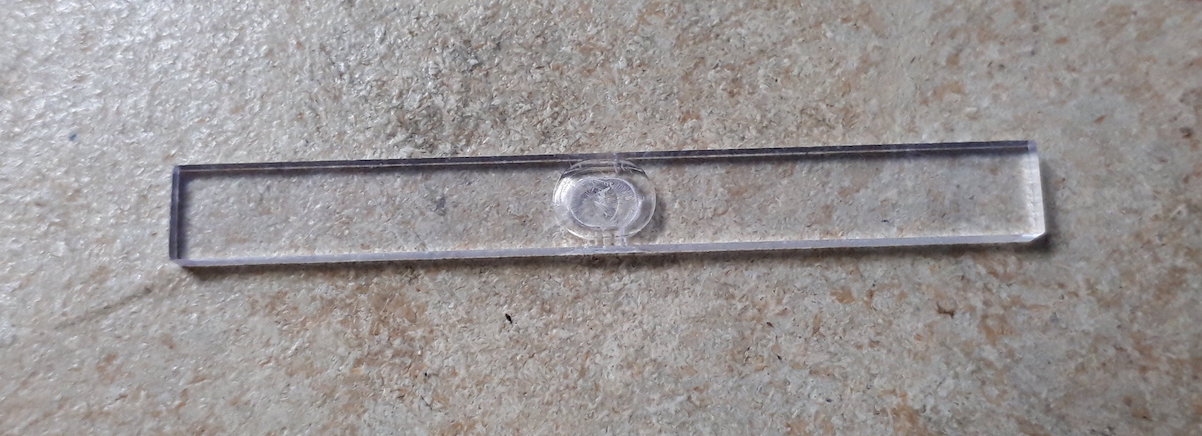}
  \caption{Scintillator strip with a dimple}\label{fig:sci_dimple_readout}
\end{figure}

First, we measured the light yield of the dimple readout scintillator for various incident positions. The measurement was performed using a scintillator strip (BC-408 made by SAINT-GOBAIN) with a dimple at the center of the strip (See \figref{fig:sci_dimple_readout}) and a $15\ \mathrm{mm}$ pitch SiPM (S12571-015P made by Hamamatsu Photonics). The SiPM is implanted in the cavity. A $^{90}\mathrm{Sr}$ checking source was used as a beta ray source; beta rays from the source were made to pass through a collimator with a diameter of $0.5\ \mathrm{mm}$. The collimator and trigger counter positions were fixed, and the strip could move along the $45\ \mathrm{mm}$ length by a moving stage.

\begin{figure}[H]
  \centering
  \includegraphics[width=0.6\linewidth]{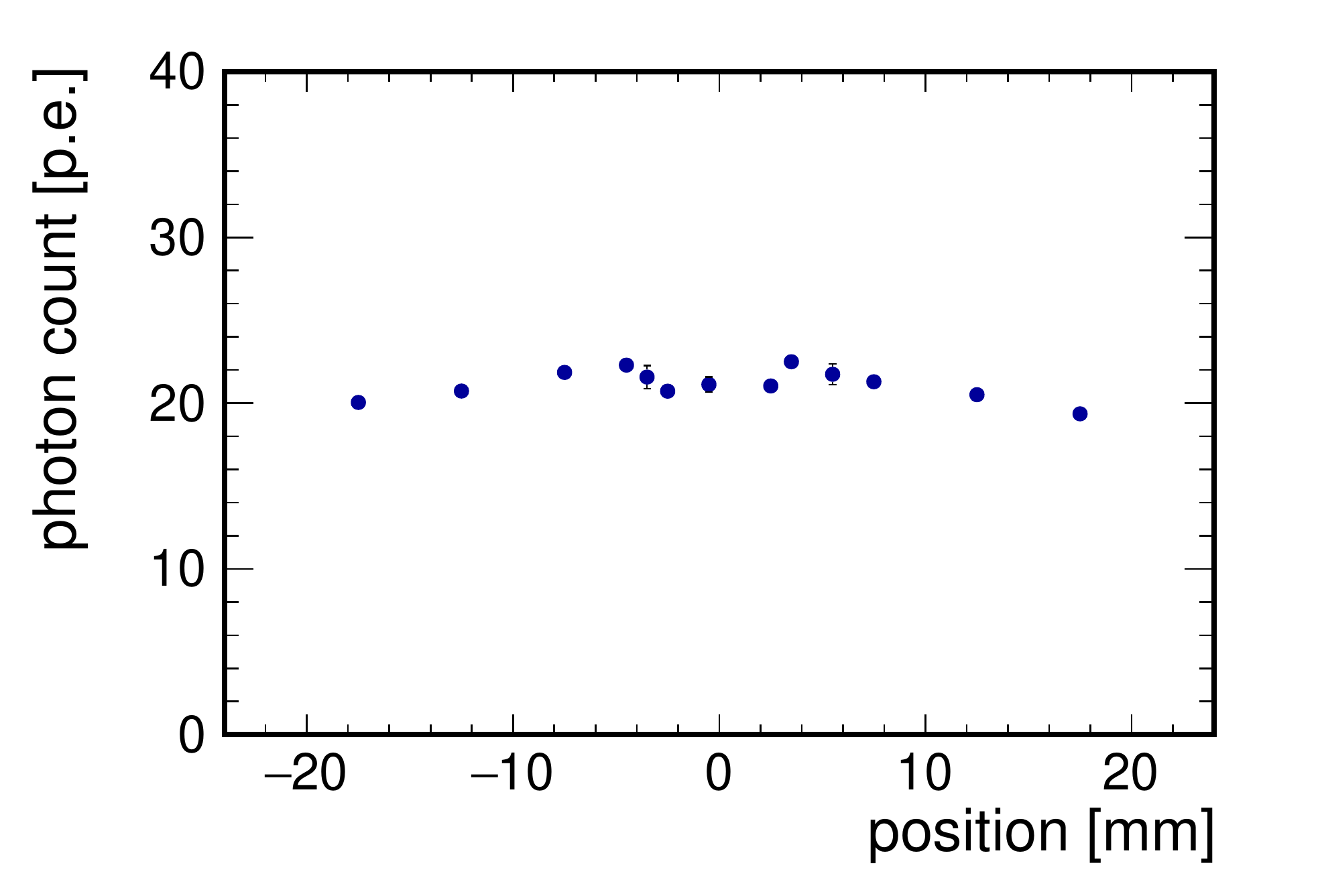}
  \caption{Result of light yield measurement for a dimple readout scintillator.}\label{fig:meas_result}
\end{figure}

\figref{fig:meas_result} shows the measurement result of position dependence. The horizontal axis is the incident position of beta ray, and the vertical axis is the average photon count at the position, which was estimated as follows. The photon count distributions from the SiPM readout were fit using a Landau distribution convoluted with a Gaussian. The most probable value of the Landau distribution was extracted as the mean photon count.

The mean light yield is about 21 p.e.\ that is about twice the photon count for the bottom readout, and this scintillator has good light yield uniformity. The width of dimple shape used in the measurement is about $5\ \mathrm{mm}$, but the width of the dip seen around the center in \figref{fig:meas_result} is about $10\ \mathrm{mm}$. This is because beta rays passed through collimator incident into the strip with spreading.

\section{Simulation}
To optimize the scintillator shape, we developed a photon tracing simulation using GEANT4 with G4OpticalPhoton class library. First, we performed the simulation under the same condition with our measurement aiming to reproduce the result of the measurement. In the simulation, electrons pass through the scintillator strip and photons are emitted inside the the scintillator strip according to the characteristic emission spectra of the BC-408 scintillator. Emitted scintillation photons move to the scintillator surface or reflector surface and then they are reflected or refracted at the surface. The photon tracking finishes when a photon gets to the photosensitive area of the SiPM or when a photon is absorbed by the reflector or scintillator.

\begin{figure}[H]
  \centering
  \includegraphics[width=0.6\linewidth]{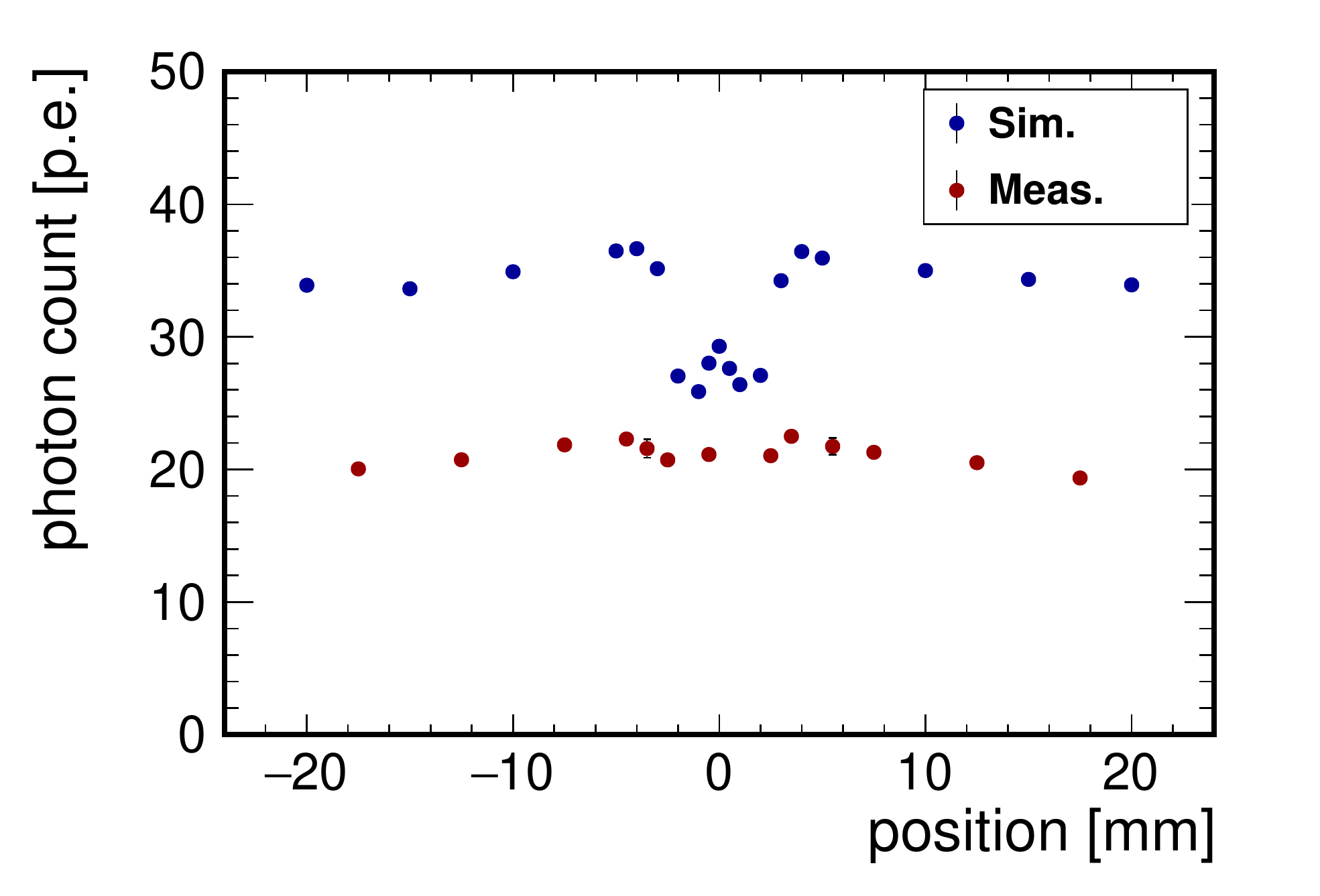}
  \caption{Result of simulation with parameters in \tbref{tb:sim_parameters}.}\label{fig:sim_detasheet}
\end{figure}

\tbref{tb:sim_parameters} shows the parameters used in the simulation. Some parameters are taken from data sheets\cite{BC_datasheet,ESR_datasheet}. \figref{fig:sim_detasheet} shows the simulation result under the condition of \tbref{tb:sim_parameters}. The Mean light yield of simulation is higher than the results of the measurement, and there is a small peak around the center in the simulation result. 
\begin{table}[H]
  \centering
  \caption{Simulation parameters}\label{tb:sim_parameters}
  \footnotesize
  \begin{tabularx}{0.9\linewidth}{C|C} \hline
    Parameter & Value \\ \hline
    size of scintillator & $45\ \mathrm{mm} \times 5\ \mathrm{mm} \times 2\ \mathrm{mm}$ \\
    depth of dimple & $0.8\ \mathrm{mm}$ \\
    refractive index & 1.58 \\
    absorption length & $380\ \mathrm{cm}$ \\
    reflectivity of ESR film & 98\% \\
    photosensitive area & $1\ \mathrm{mm} \times 1\ \mathrm{mm}$ \\
    depth of collimator & $3\ \mathrm{mm}$ \\
    diameter of collimator & $0.5\ \mathrm{mm}$ \\ \hline
  \end{tabularx}
\end{table}

\begin{figure}[H]
  \centering
  \begin{minipage}[c]{0.49\linewidth}
    \includegraphics[width=\linewidth]{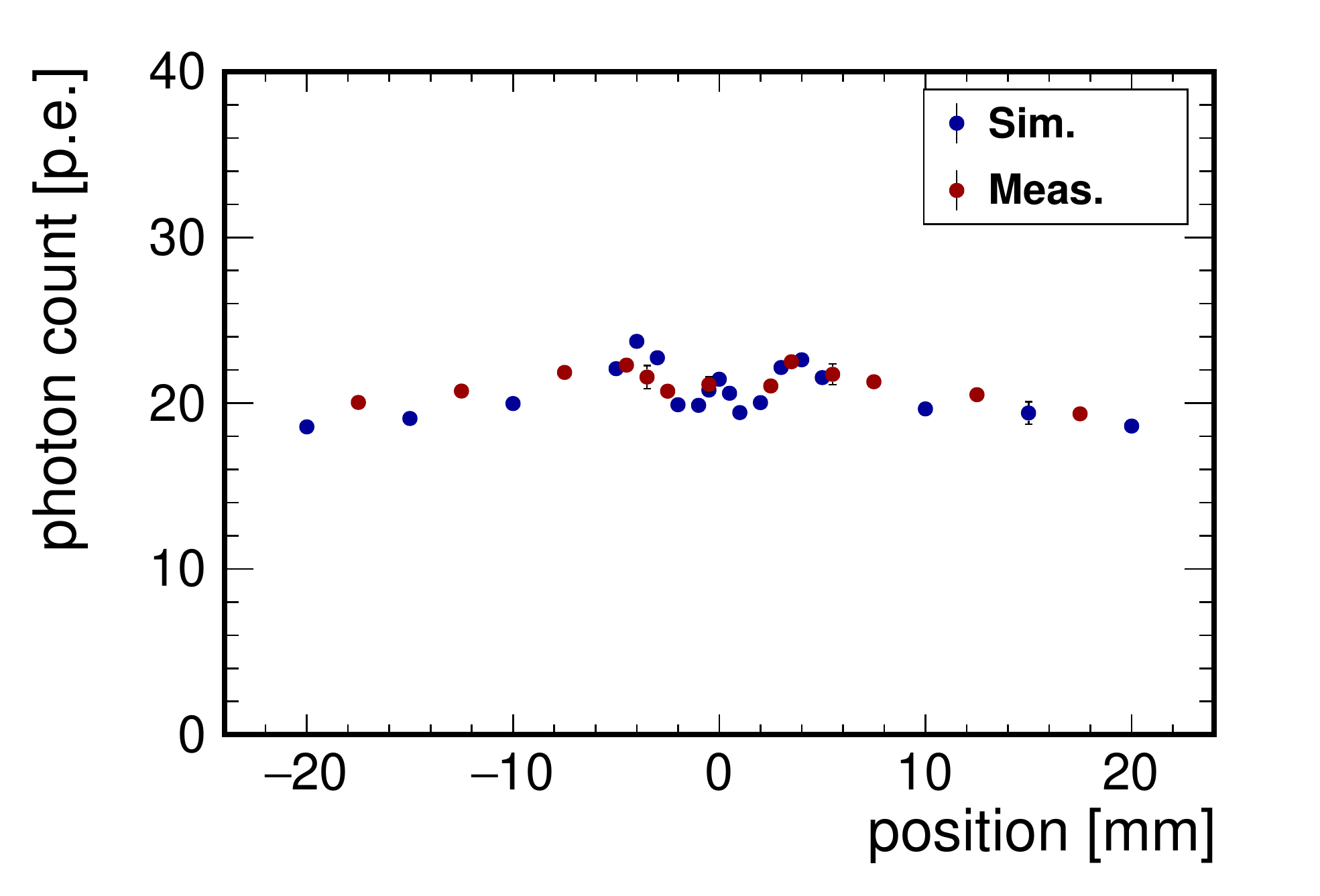} 
  \end{minipage}
  \begin{minipage}[c]{0.49\linewidth}
    \includegraphics[width=\linewidth]{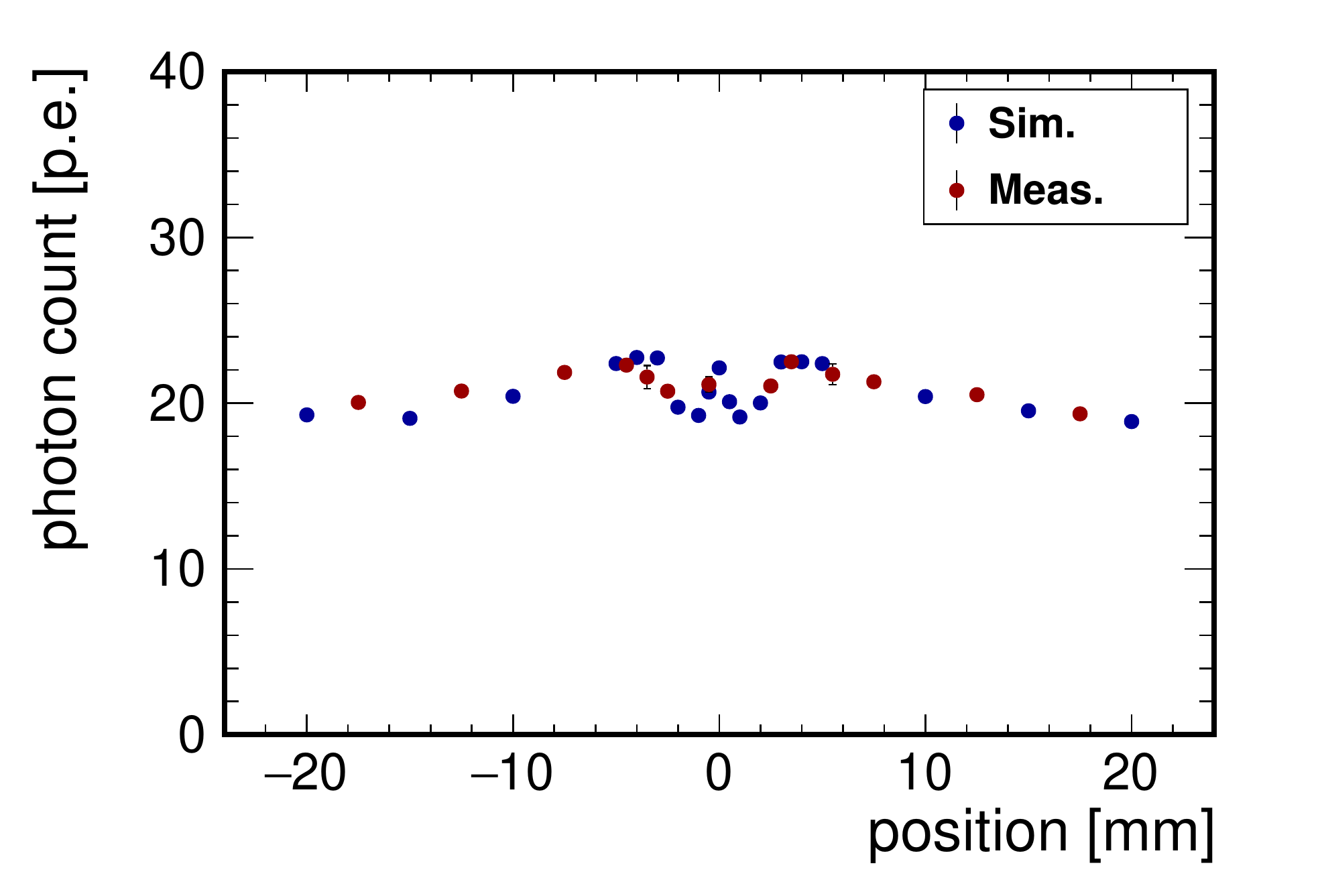}
  \end{minipage}
  \caption{Results of simulation for different parameters. (left) reflectivity: 95\% and absorption length: $200\ \mathrm{cm}$. (right) reflectivity: 96\% and absorption length: $100\ \mathrm{cm}$.}\label{fig:sim_diff_parameter}
\end{figure}

In the previous simulation, we inputted some optical parameters as fixed value. Parameters about optical properties such as the reflectivity of the reflector film and bulk light attenuation length of the scintillator, however, should depend on the wavelength of scintillation photon. So these values would be deviated from the value in the data sheets described as fixed values. By tuning these parameters, the measurement result can be reproduced as shown in \figref{fig:sim_diff_parameter} (left). On the other hand, different parameter values also reproduce the result as in \figref{fig:sim_diff_parameter} (right).

\begin{figure}[H]
  \centering
  \begin{minipage}[c]{0.49\linewidth}
    \includegraphics[width=\linewidth]{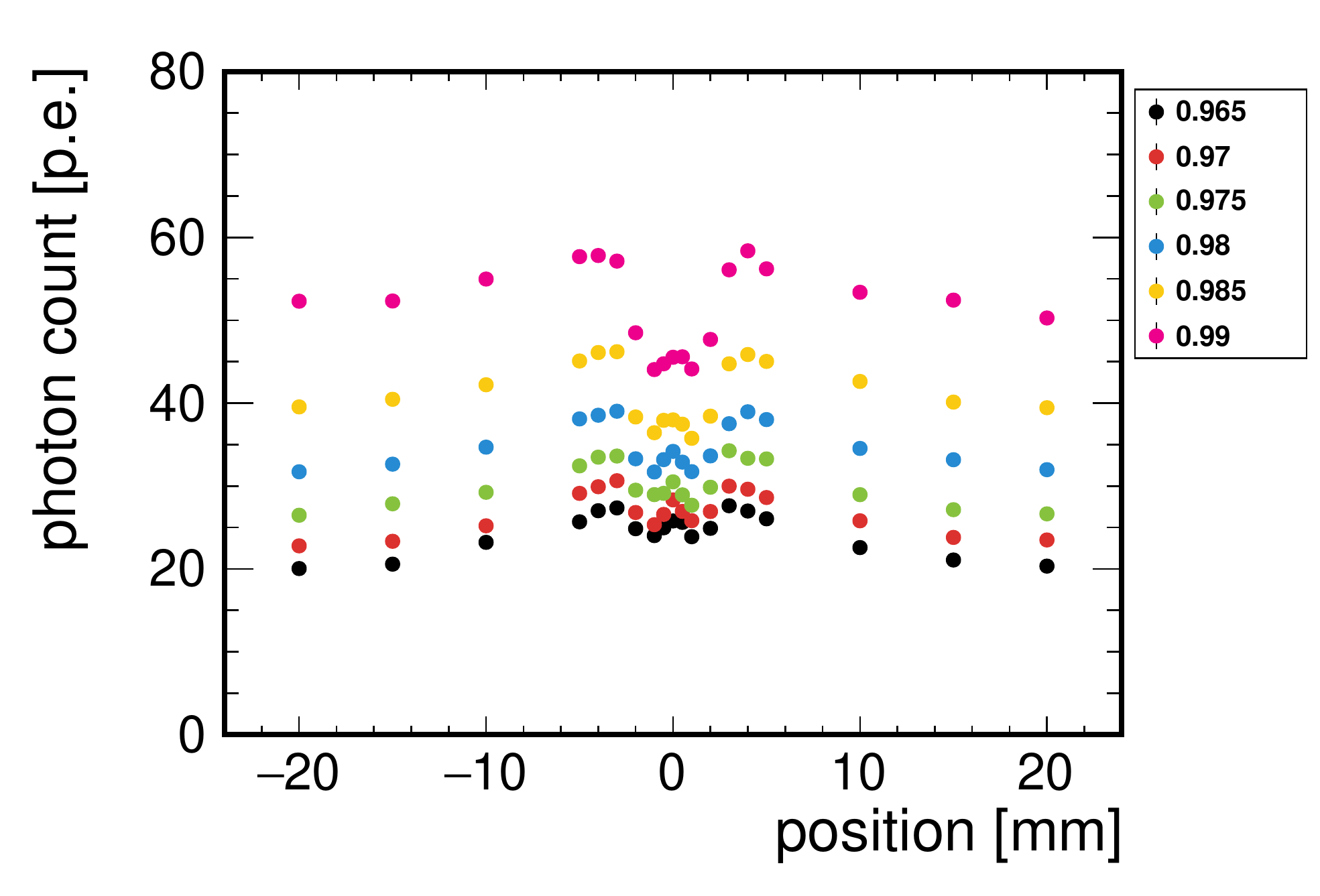} 
  \end{minipage}
  \begin{minipage}[c]{0.49\linewidth}
    \includegraphics[width=\linewidth]{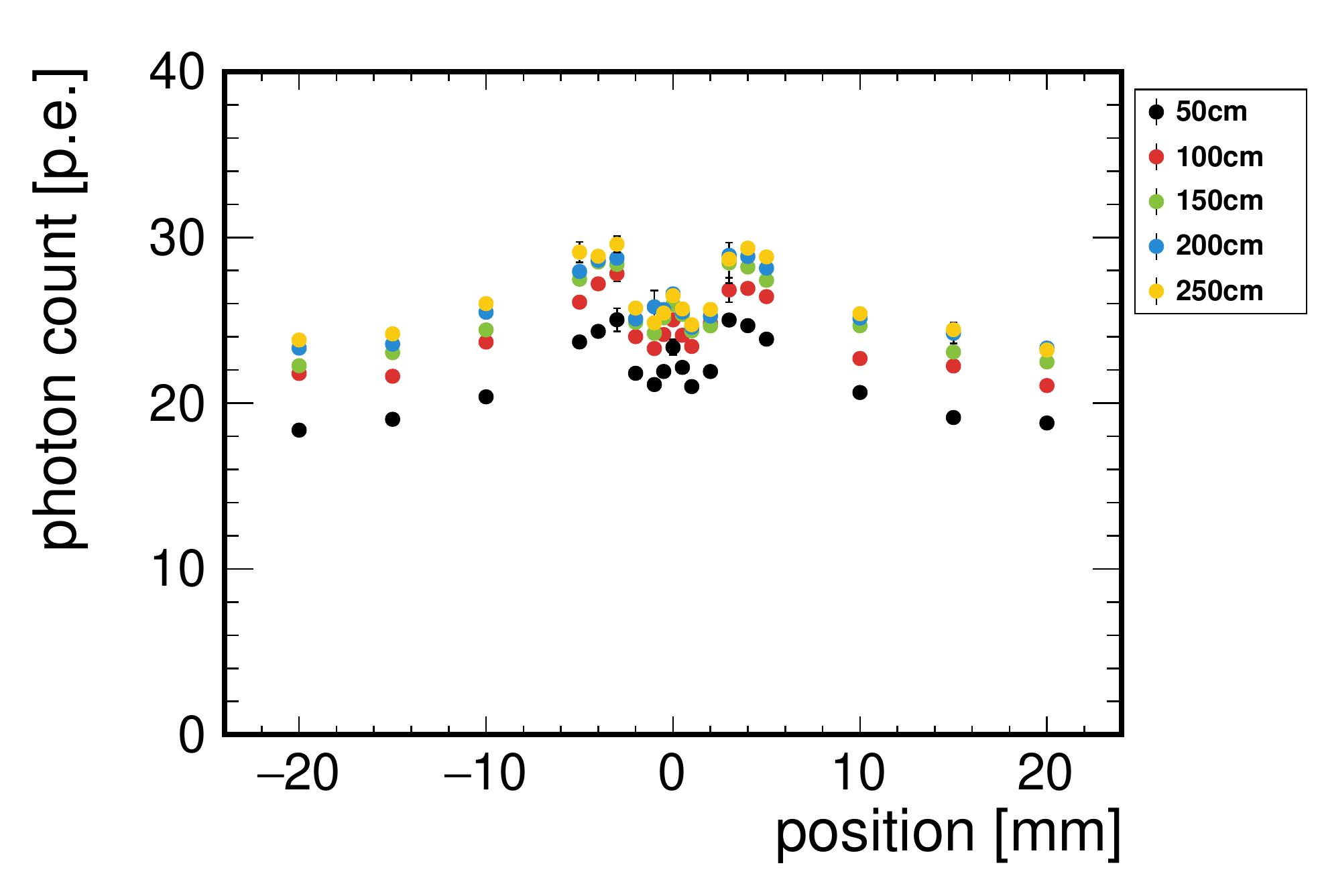}
  \end{minipage}
  \caption{Results of simulation for different optical properties. (left) Light yield for different reflectivities. (right) Light yield for different light absorption lengths.}\label{fig:sim_param_scan}
\end{figure}

\figref{fig:sim_diff_parameter} suggests that optical properties such as the reflectivity, the absorption length, and the refractive index cannot be determined uniquely by simulation. Then, we investigated the dependency of optical parameters using simulation. \figref{fig:sim_param_scan} (left) shows the results for different reflectivities of the reflector film and \figref{fig:sim_param_scan} (right) shows the results for different light attenuation lengths. These figures show a small variation of these values has a large effect on the light yield. For example, the light yield is reduced by half when changing the reflectivity by 1\%. So it is necessary to accurately determine the parameters about optical properties of the scintillator and the reflector film. As these properties depend on the photon wavelength, their values must be measured for several wavelengths.

\begin{figure}[H]
  \centering
  \begin{minipage}[c]{0.49\linewidth}
    \includegraphics[width=\linewidth]{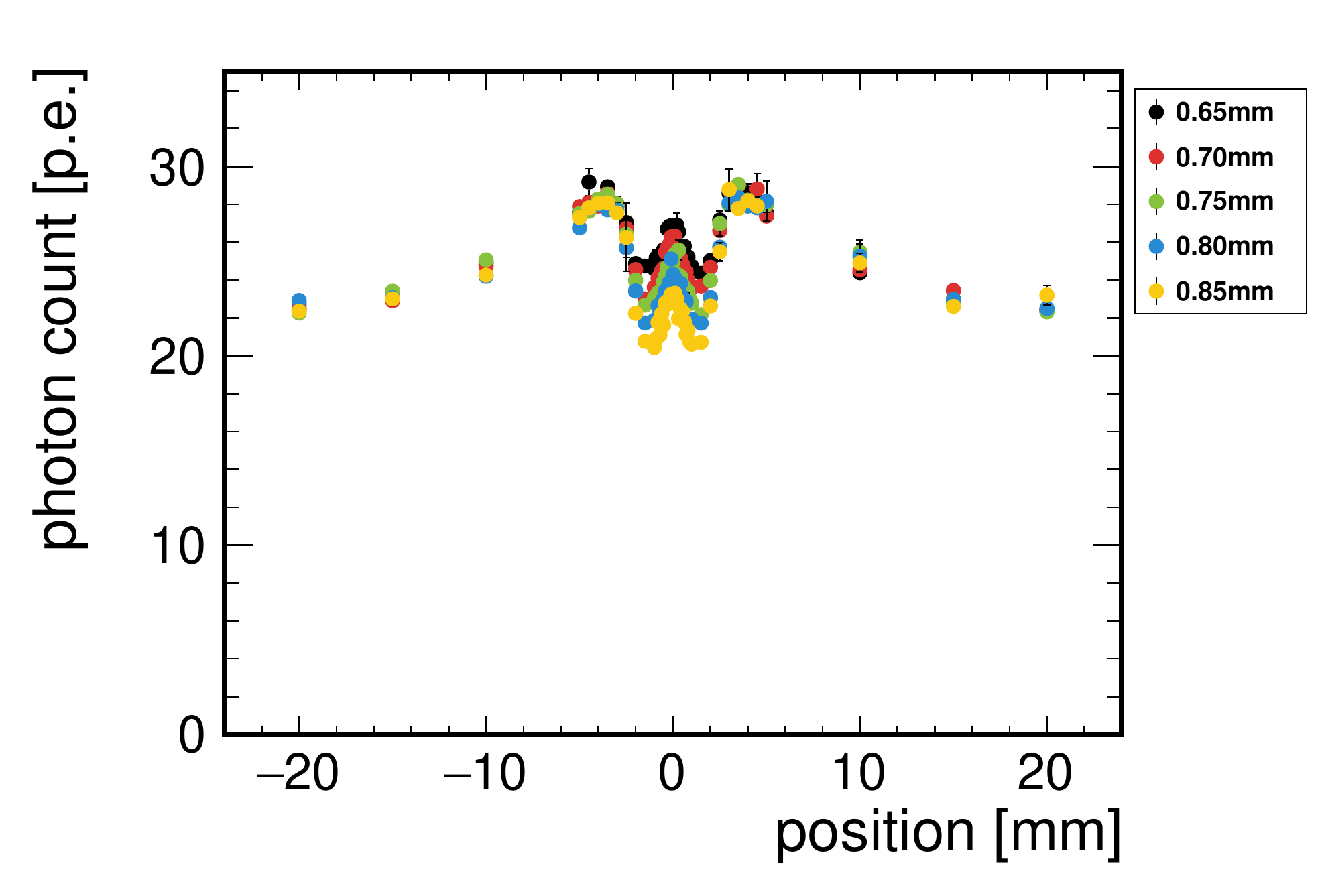} 
  \end{minipage}
  \begin{minipage}[c]{0.49\linewidth}
    \includegraphics[width=\linewidth]{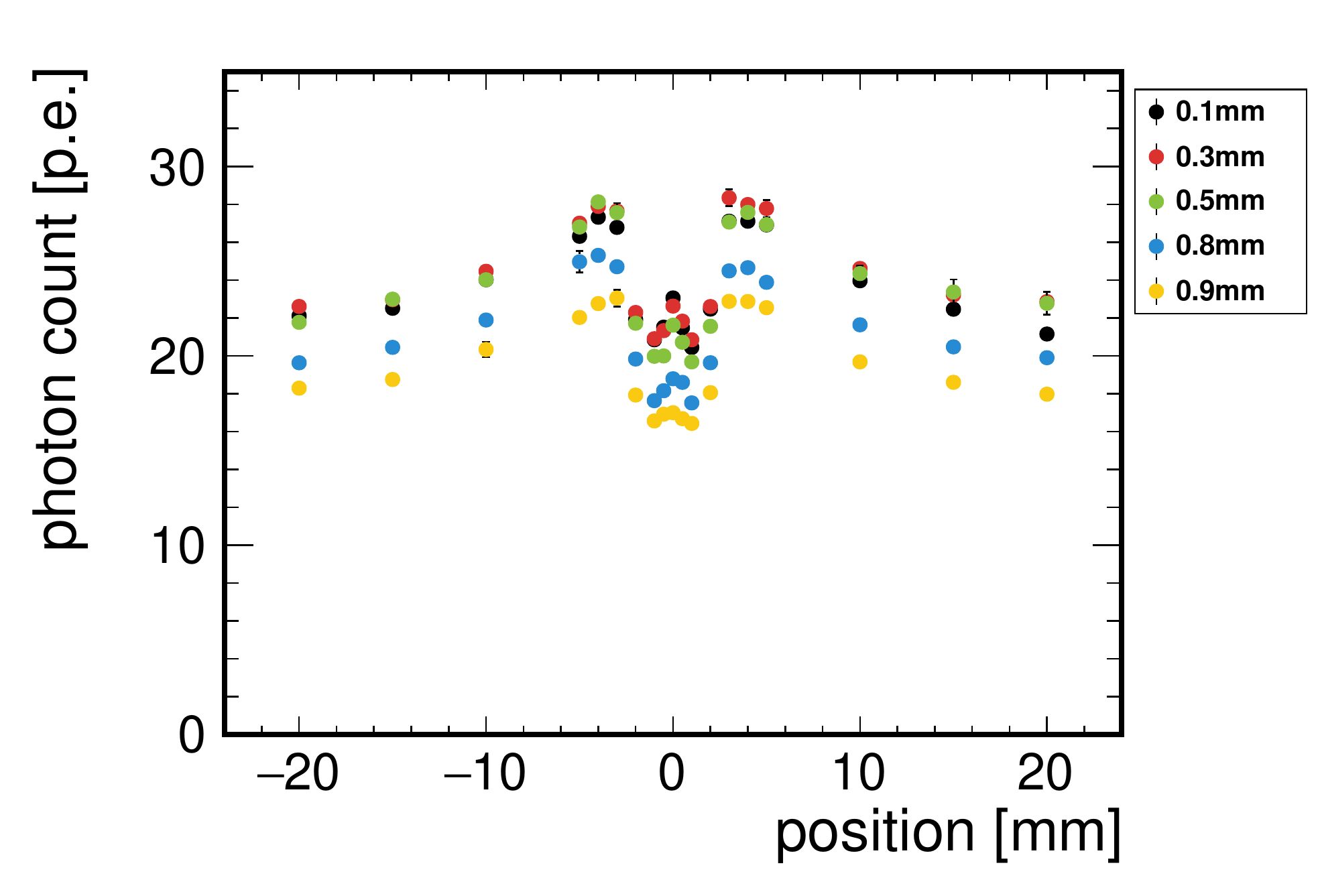}
  \end{minipage}
  \caption{Results of simulation for the different geometry. (left) Light yield for different depths of the dimple. (right) Light yield for different distances from the dimple surface to the SiPM.}\label{fig:sim_geom_scan}
\end{figure}

Next, we investigated the effects of the dimple size and the position of the SiPM on the detected light yield. \figref{fig:sim_geom_scan} (left) shows the results for different depths of the dimple and \figref{fig:sim_geom_scan} (right) shows the results for different distances from the dimple bottom surface to the top of the SiPM. The light yield around the dimple decreases by decreasing the depth of the dimple, and the entire light yield decreases by increasing the distance from the dimple surface to the SiPM. In the dimple readout method, scintillation photons can enter into the package of the SiPM from the side. This is important for optimizing the scintillator shape.

\section{Summary}
We are developing scintillator-based electromagnetic calorimeter for Higgs factories. We confirmed that the dimple readout scintillator has good light yield and good uniformity, and we are making simulation code for optimizing the scintillator shape. Simulation software was developed that can reproduce the behavior of our measurement. Some optical parameters such as the reflectance of the reflector film and the bulk light absorption length of the scintillator strip have large effect on the light yield detected in the strip. To match the result of simulation and measurement, we need dedicated measurements about optical properties.

\section*{Acknowledgements}
We thank CALICE-Asia group members and researchers of IHEP and USTC for various advices on this study.